\documentclass[]{iopart}
\usepackage{graphicx}
\usepackage{iopams}
\begin{document}

\title{Cooperative light scattering from helical-phase-imprinted atomic rings}

\author{H. H. Jen}
\address{Institute of Physics, Academia Sinica, Taipei 11529, Taiwan}
\ead{sappyjen@gmail.com}

\author{M.-S. Chang}
\address{Institute of Atomic and Molecular Sciences, Academia Sinica, Taipei 10617, Taiwan}
\author{Y.-C. Chen}
\address{Institute of Atomic and Molecular Sciences, Academia Sinica, Taipei 10617, Taiwan}
\renewcommand{\k}{\mathbf{k}}
\renewcommand{\r}{\mathbf{r}}
\newcommand{\parallelsum}{\mathbin{\!/\mkern-5mu/\!}}
\def\p{\mathbf{p}} 
\def\R{\mathbf{R}}
\def\bea{\begin{eqnarray}}
\def\eea{\end{eqnarray}}
\begin{abstract} 
We theoretically investigate the light scattering of the super- and subradiant states which can be prepared by the excitation of a single photon which carries an orbital angular momentum (OAM).\ With this helical phase imprinted on the stacked ring of atomic arrays, the subradiant modes show directional side scattering in the far-field, allowing for light collimation and quantum storage of light with OAM.\ For the excitations with linear polarizations, we find a discrete $C_4$ rotational symmetry in scattering for the number of atoms $N$ $=$ $4n $ with integers $n$, while for circular polarizations with arbitrary $N$, the azimuthal and $C_N$ symmetries emerge for the super- and subradiant modes respectively.\ When the radial and azimuthal polarizations are considered, a mode shift can happen in the scattering pattern.\ The forward scattering of the superradiant modes can be enhanced as we stack up the rings along the excitation direction, and for the subradiant modes, we find the narrowing effects on the scattering in the azimuthal and the polar angles when more concentric rings are added in the radial direction.\ By designing the atomic spatial distributions and excitation polarizations, helical-phase-imprinted subradiant states can tailor and modify the radiation properties, which is detectable in the directional super- and subradiant emissions and is potentially useful in quantum information manipulations. 
\end{abstract}
\maketitle
\section{Introduction}

Controlled strong light-matter interactions in quantum optical systems for efficient generation, storage, and manipulation of quantum correlations \cite{Hammerer2010} is essential for establishing robust long-distance quantum entanglement for quantum communication \cite{Chaneliere2006, Radnaev2010, Jen2010} and quantum network \cite{Kimble2008}. This has also expedited the development of quantum memory and quantum repeater \cite{Duan2001}, which respectively stores and relays entanglement. Quantum correlation in neutral atomic system is often arisen through spontaneous emissions, in which the atomic ensemble collectively and spontaneously emits a photon following an atomic excitation. This serves an elementary mechanism to entangle the atomic states with discrete states of light, such as polarizations \cite{Clauser1969, Aspect1981, Kwiat1995}. This bipartite entanglement can also be generated in the biphoton states in spatial modes, or energy-time \cite{Braunstein2005}, using either parametric down conversion from nonlinear crystals \cite{Law2000,Parker2000} or cascade atomic configurations \cite{Jen2012-2,Jen2016a,Jen2016b,Jen2017_cascade}. By independently entangle more than one degrees of freedom, one then achieves hyper-entanglement which allows increasing the information capacity of the carriers.

To increase the efficiency of light-matter interaction, directional spontaneous emissions are enhanced in optically thick atomic ensembles \cite{Bromley2016,Zhu2016,Shahmoon2017} through superradiance \cite{Dicke1954, Gross1982}, utilizing the resonant and pairwise dipole-dipole interactions \cite{Stephen1964,Lehmberg1970} among the atoms in the dissipation process. This collective light-matter interaction also results in a frequency shift \cite{Friedberg1973, Scully2009,Rohlsberger2010,Keaveney2012,Meir2014,Pellegrino2014,Jen2015,Jennewein2016,Jenkins2016} and is responsible for subradiant radiations \cite{Guerin2016} as an afterglow of superradiance \cite{Mazets2007}. In the context of quantum memory, subradiant states are candidate systems for storing photons and can be actively prepared in a dense atomic medium \cite{Scully2015,Jen2016_SR,Jen2016_SR2,Sutherland2016,Bettles2016,Jen2017_MP}, through selective radiance by controlling the positions of an array of atoms or metamolecules \cite{Garcia2017, Facchinetti2016, Jenkins2017}, collective antiresonances from the subradiant arrays in a cavity \cite{Plankensteiner2017}, or through creating quantum optical analogs of topological states in two-dimensional atomic arrays \cite{Perczel2017}.  The light scattering from the subradiant states is also under active investigations recently \cite{Guerin2016, Shahmoon2017}, but a systematic and detailed study on the subradiant modes is still lacking.

The rapid development on precisely positioning single atoms utilizing photonic crystal waveguide \cite{Goban2015}, optical microtraps \cite{Barredo2016, Endres2016}, or creating an array of artificial atoms in solid-state nanophotonic platforms \cite{Sipahigil2016} has further enabled fabrication of atomic ensembles with arbitrary spatial distributions beyond the diffraction limit of the excitation field, thus offers new opportunities to explore super- and subradiant modes, and opens up a new avenue for tailoring and modifying the quantum states of light and matter. In this paper, we propose to prepare the phase-imprinted single-photon subradiant states in the stacked ring arrays of atoms, in which light with orbital angular momentum (OAM) \cite{Arnaut2000, Mair2001, Molina2007, Dada2011, Fickler2012} interacts with the atoms, and the helical-phase-imprinted (HPI) subradiant states can be prepared upon absorption. The HPI subradiant states thus serves as good candidates for storing light quanta with OAM \cite{Nicolas2014, Ding2015, Zhou2015}. We investigate the light scattering out of these subradiant states for the cases of a few atoms, a single ring, and stacked rings. We also study the effect of uniform and spatially-dependent light polarizations on the scattering patterns.

\section{Helical-phase-imprinted subradiant states}

When a near-resonant single photon is absorbed by an ensemble of $N$ two-level atoms, a symmetric state is formed, 
\bea
|\Phi_N\rangle=\frac{1}{\sqrt{N}}\sum_{\mu=1}^N e^{i\k_L\cdot\r_\mu}|e\rangle_\mu|g\rangle^{\otimes(N-1)},
\eea
where each of two-level atoms can be promoted to the excited state $|e\rangle$ from the ground state $|g\rangle$ with an equal probability $N^{-1}$ and a position-dependent traveling phase $e^{i\k_L\cdot\r_\mu}$ given by the plane-wave excitation field. This symmetric state can be superradiant when the inter-atomic distance is much less than the resonant wavelength $\lambda_a$.\ Since the complete Hilbert space of single excitation involves $N$ possible constructions of the bare states $|e\rangle_\mu|g\rangle^{(N-1)}$ $\equiv$ $|\psi_\mu\rangle$, then the remaining $N-1$ nonsymmetric states can be either super- or subradiant, depending on the atomic distributions.\ To systematically study and access those states, we have proposed to use a phase-imprinting method \cite{Jen2016_SR,Jen2016_SR2,Jen2017_MP} on a one-dimensional atomic array, which prepares the system into a De Moivre state:
\bea
|\Phi_m\rangle=\frac{1}{\sqrt{N}}\sum_{\mu=1}^N e^{i\k_L\cdot\r_\mu}e^{i\frac{2m\pi}{N}(\mu-1)}|\psi_\mu\rangle, \label{DM}
\eea 
with $m$ $\in$ $[1,N]$, whose orthogonality is guaranteed by De Moivre¡¦s theorem.\ This phase imprinting method dynamically controls the linearly increased atomic phases either by a gradient Zeeman field or a gradient Stark field pulse. Note that this specific construction of Hilbert space is not unique, and in general there are infinite ways to create single excitation space.

\begin{figure}[t]
\centering
\includegraphics[width=10.0cm,height=7.5cm]{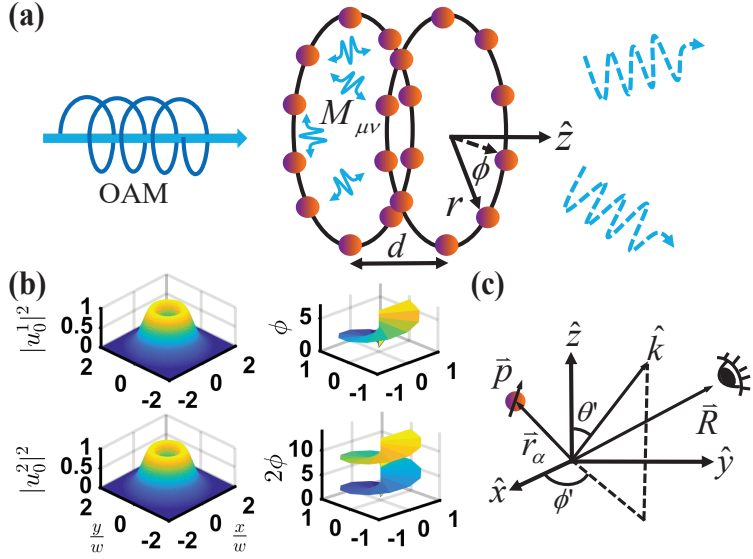}
\caption{Schematic helical-phase-imprinted state preparation and far-field detection.\ (a) A single-photon source with orbital angular momentum (OAM) is absorbed by the atoms sitting on the stacked rings along $\hat z$ (two ring arrays are shown here for illustration).\ The atomic system is then prepared into one of the super- or subradiant states, depending on OAM of light.\ In the dissipation process, the resonant dipole-dipole interaction $M_{\mu\nu}$ couples any two atoms on the stacked rings, and scatters the light collectively depending on the ring geometry of radius $r$, $\phi$, and inter-ring distance $d$.\ (b) Typical light intensity (normalized) with OAM of Laguerre-Gaussian modes $u_0^l(r,\phi)$ and associated helical phases $e^{il\phi}$, for some beam waist radius $w$.\ (c) A far-field observer at $\vec R$ sees the scattered light from a dipole $\vec p$ at $\vec r_{\alpha}$ in $4\pi$ solid angle of mode $\hat k$ characterized by $\theta'$ and $\phi'$.}\label{fig1}
\end{figure}

While this method practically demands a large field gradient or long interaction time when the atomic array is short and/or the inter-atomic separation is small, by deforming the atomic array into a ring, this linearly increasing phase can be easily and exactly imprinted by light with a quantized orbital angular momentum (OAM), $l\hbar$, without needing an external field gradient. A Laguerre-Gaussian (LG) beam, $u^l_0$, carries an OAM of $l\hbar$, and its wave front acquires a phase $e^{il\phi}$ along the azimuthal direction, where the azimuthal angle $\phi=[0, 2\pi)$ \cite{Arnaut2000, Mair2001}. For $N$ atoms sitting on a single ring with a constant separation between their nearest neighbors, the light propagating along the axis of the ring imprints the phase of $\phi= 2\pi l(\mu-1)/N$ on the atoms. A $u^l_0$ photon absorbed by the ring array thus forms exactly the state of $|\Phi_{m=l}\rangle$ of equation (\ref{DM}). In Figure \ref{fig1}, we show two stacked rings for an illustration of preparing such helical-phase-imprinted (HPI) states with a $u^l_0$ photon, and the far-field observation of these states.\ For the multiply-stacked rings along $\hat z$, HPI states in general can be expressed as

\bea
|\Phi_l\rangle_{\rm HPI}=\frac{1}{\sqrt{N}}\sum_{\mu_z=1}^{N_z}\sum_{\mu_\phi=1}^{N_\phi}e^{i\k_L\cdot\r_\mu}e^{i\frac{2l\pi}{N_\phi}(\mu_\phi-1)}|\psi_\mu\rangle, \label{DM2}
\eea
where $N$ $=$ $N_z N_\phi$ for a number of $N_z$ stacked rings, with $N_\phi$ atoms in each ring.\ The atomic position index $\mu$ is implicitly $(\mu_z$$-$$1)$$N_\phi$ $+$ $\mu_\phi$, which labels the traveling phase by the excitation field on $|\psi_\mu\rangle$.\ The multiply-stacked rings allow a large optical depth and strong light-matter interactions, and thus increase the absorption efficiency.\ For the stacked rings along $\hat r$, forming a concentric structure in a two-dimensional plane, we can substitute $N_z$ and $\mu_z$ with $N_r$ and $\mu_r$ respectively in equation (\ref{DM2}).\ Other possible HPI states can be prepared in a cylindrical shell with a chirality or even in torus-like shape, making our scheme a versatile platform to manipulate and engineer the many-body subradiant states.  

\section{Light scattering from HPI subradiant states}

In this section, we consider the resonant dipole-dipole interaction in the dissipation process of the HPI states and investigate their far-field emission patterns, which is derived in Appendix B, for various atomic ring structures.

\subsection{Two-atom case.}

\begin{figure}[t]
\centering
\includegraphics[width=10.0cm,height=5cm]{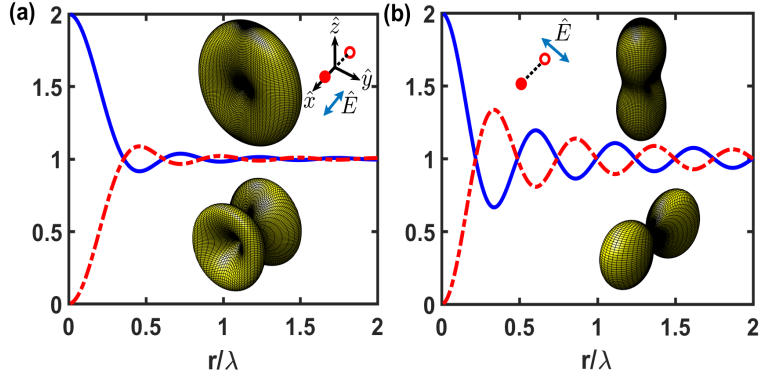}
\caption{Eigenvalues and far-field property $\Omega_f(\theta,\phi)$ for two atoms sitting on the $\hat x$ axis.\ We show the real part of the rescaled eigenvalues, $-\lambda_{1,2}/(\Gamma/2)$, in the case of two atoms separating by $2r$, for the eigenstates of $l$ $=$ $1$ (red dashed line, subradiant) and $2$ (blue solid line, superradiant) with (a) $\hat x$ and (b) $\hat y$ polarized light excitations $\hat E$.\ Specific three-dimensional plots of $\Omega_f(\theta,\phi)$ at $r/\lambda$ $=$ $0.15$ for the super- and subradiant states are illustrated respectively in the upper and lower parts of the graphs. Empty and filled circles indicate the ground and excited atoms, one of the bare states $|\psi_\mu\rangle$.}\label{fig2}
\end{figure}

We first analyze the case of two atoms sitting on a ring with a radius $r$ and excited by single photon carrying an OAM, $l\hbar$. Define the far-field property $\Omega_f(\theta,\phi)$ $\equiv$ $\langle\vec E^*(\R,t)\vec E(\R,t)\rangle/I_0(t)$, we use equation (\ref{far}) in Appendix B for two atoms on the $x$-axis with an $\hat x$ polarized excitation, and we obtain 
\bea
\Omega_f(\theta,\phi)=(1-\sin^2\theta\cos^2\phi)\left[2+2\cos(2|\k_L|r\sin\theta\cos\phi+l\pi)\right],\label{far2}
\eea
which corresponds to the case shown in figure \ref{fig2}(a), where two atoms are aligned parallel to the polarization of the excitation field. For the case where two atoms are aligned perpendicular to the polarization which is shown in figure \ref{fig2}(b), the factor in equation (\ref{far2}), $1-\sin^2\theta\cos^2\phi$, is replaced by $\sin^2\theta\sin^2\phi$.\ Note that  $|\k_R|$ $=$ $|\k_L|$, and different light polarizations result in different coupling strengths in equations (\ref{F}) and (\ref{G}), and thus $I_0(t)$, due to different eigenvalues $\lambda_m$ (see Appendix A).\ When $r$ $\rightarrow$ $0$, $\Omega_f(\theta,\phi)$ $\propto$ $[1+(-1)^l]$, which indicates that the excitation beam with odd OAM is not scattered at all in this extreme limit. According to equation (\ref{DM2}), this specific HPI state is given as
\bea
|\Phi_l\rangle_{\rm HPI}=\frac{1}{\sqrt{2}}\left(e^{i\k_L\cdot\r_1}|\psi_1\rangle+e^{i\k_L\cdot\r
_2}e^{il\pi}|\psi_2\rangle\right),\label{2atom}
\eea
which is a superradiant (subradiant) state for even (odd) $l$. The superradiant intensity for single photon scattering in the forward direction has a maximal $\Omega_f(\theta=0)$ $=$ $2$ which is proportional to $N^2/2$, as that in the half-excited spin models \cite{Dicke1954}.\ For single spin excitation in the  $N$ spin-$1/2$ model, the photon emission intensity is proportional to $(l_m$ $+$ $m)(l_m$ $-$ $m$ $+$ $1)$ $=$ $N$ in the Dicke's eigenstates with a total quantized angular momentum $l_m\hbar = N\hbar/2$ and magnetization $m$ $\equiv$ $(N_\uparrow$ $-$ $N_\downarrow)/2$ $=$ $1$ $-$ $N/2$ \cite{Dicke1954}.  

In Figure \ref{fig2}, we show the eigenvalues and far-field property $\Omega_f(\theta,\phi)$ for two atoms separating by $2r$.\ The eigenvalues can be solved analytically from the coupling matrix $\hat M$ introduced in Appendix A, which are
\bea 
\lambda_{1,2}=-\frac{\Gamma}{2}\pm \left[\frac{F_{12}(\xi)-i2G_{12}(\xi)}{2}\right],
\eea
with $\xi$ $=$ $2|\k_L|r$.\ The rescaled real part of the eigenvalues in Figure \ref{fig2}, which are decay constants, approach $2$ and $0$ as $r$ $\rightarrow$ $0$, representing the super- and subradiant modes of the radiation.\ For larger $r$, the eigenvalues asymptotically converge to $1$, corresponding to the regime of non-interacting (independent) emitters.\ Specific far-field property is chosen at $\xi$ $=$ $0.3\pi$, which shows a forward-backward scattering along the propagation direction, $\hat z$, of the excitation field and a side scattering respectively for the super- ($l$ $=$ $2$) and subradiant ($l$ $=$ $1$) modes.\ Note that the $\hat x$ and $\hat y$ polarized excitations correspond to the head-to-tail and parallel polarization configurations respectively.\ Therefore, the former (latter) shows no scattered light at all in the direction of $\hat x$ ($\hat y$), and this can be also seen from equation (\ref{far2}) which vanishes at $\theta$ $=$ $\pi/2$ and $\phi$ $=$ $0$ ($\pi/2$).\ The head-to-tail polarizations also show strong scattering in the superradiant mode in the transverse direction to the polarization orientation, in contrast to the parallel polarizations which have a suppressed scattering rate along the $\hat x$ axis (about $0.35$ times of the maximal scattering).\ This reflects the destructive light interference between the parallel polarizations and can be seen as phase slip in the scattered light.\ This destructive interference goes away as $r$ $\rightarrow$ $0$, where $\Omega_f(\theta,\phi)$ of these two polarization configurations restores the rotational symmetry of $\phi$ $\rightarrow$ $\phi$ $+$ $\pi/2$.

On the contrary, the subradiant modes preserve the scattering patterns in Figure \ref{fig2} at $r/\lambda$ 
$\lesssim$ $0.35$ and $0.25$ for head-to-tail and parallel polarizations respectively.\ This range of $r$ can be estimated by $2|\k_L| r\sin\theta\cos\phi$ $\approx$ $\pi/2$ in equation (\ref{far2}), which indicates the phase slip of half of $l\pi$ ($l$ $=$ $1$ for the subradiant mode).\ The angles can be chosen as $\phi$ $=$ $0$ and $\theta$ $=$ $\pi/4$ or $0$ at the maximal scattering of the head-to-tail or parallel polarizations in the small $r$ limit.\ This estimation also reflects on the qualitative change of the eigenvalues in Figure \ref{fig2}, which start to oscillate around $\lambda_{1,2}$ $=$ $-\Gamma/2$.\ As $r$ increases and passes the estimated range, a side scattering also appears for the superradiant modes, and the directionality of the scattering disappears.\ In the range where $2r$ $\gtrsim$ $\lambda$, the clear phenomena of super- and subradiance disappear. 

\subsection{Single ring.} 

For the geometry of $N$ atoms sitting on a single ring with equal arc lengths, it is equivalent to  an $N$-sided regular polygon.\ When $N \gg 1$,  the far-field scattering pattern from atoms forming a regular polygon approaches that of a ring.\ Before we investigate the scattering of the many-body subradiant states in a single ring with a large $N$, we first study the case of three and four atoms, which form regular triangle and square respectively.\ 

\begin{figure}[t]
\centering
\includegraphics[width=10.0cm,height=5cm]{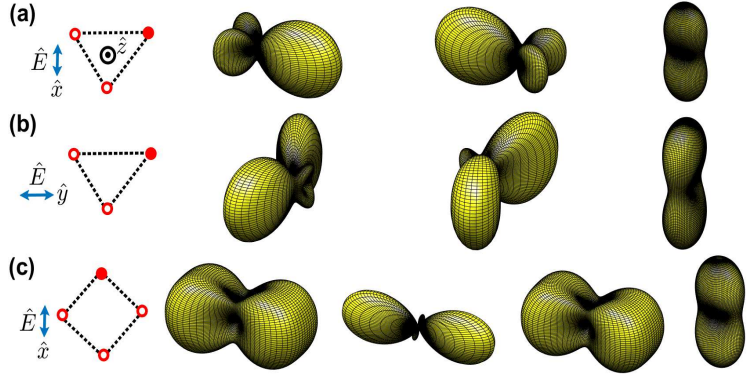}
\caption{The far-field $\Omega_f(\theta,\phi)$ for three and four atoms sitting on a single ring.\ The atoms in top view form an (a) $\hat x$ and (b) $\hat y$ polarized triangular, and (c) $\hat x$ polarized square structures with the modes of $l$ $=$ $1-N$ respectively, where we choose $r/\lambda$ $=$ $0.2$.\ Subradiant modes ($l$ $=$ $1$ to $N-1$) and superradiant modes ($l$ $=$ $N$) show directional side and forward-backward scatterings respectively.\ Note that $\Omega_f(\theta,\phi)$ of $\hat y$ polarized square preserves the $C_4$ rotational symmetry to the case of (c), and the viewing angles are the same as Figure \ref{fig2}. Again empty and filled circles represent the ground and excited atoms, which displays one of the bare states $|\psi_\mu\rangle$.}\label{fig3}
\end{figure}

In Figures \ref{fig3}(a) and \ref{fig3}(b), we show two polarization configurations of three atoms in a triangular, which are excited by $\hat x$ and $\hat y$ polarized light respectively.\ The far-field scatterings $\Omega_f(\theta,\phi)$ of various HPI states of $l$ $=$ $1-3$ are plotted horizontally.\ As expected, the superradiant modes of $l$ $=$ $3$ show directional forward-backward scatterings along $\hat z$.\ In contrast, for the subradiant modes of $l$ $=$ $1$ and $2$, a side scattering shows up but its mirrored counterpart is suppressed with respective to the $\hat x-\hat z$ plane.\ This asymmetry is due to the finite phase slips between the atoms, and the mirror symmetries to the $\hat x-\hat z$ and $\hat y-\hat z$ planes can be restored as $r$ $\rightarrow$ $0$, where the maximal scatterings reside on the $\hat y$ and $\hat x$-axis respectively for \ref{fig3}(a) and \ref{fig3}(b).\ Unlike the two-atom case, the rotational symmetry of $\Omega_f(\theta,\phi)$ for both the super- and subradiant modes, that is $\phi$ $\rightarrow$ $\phi$ $+$ $\pi/2$, is retrieved as $r$ $\rightarrow$ $0$.\ This reflects the role of the third atom which smears the pure parallel or head-to-tail polarization configurations in the two-atom case.\ In Figure \ref{fig3}(c), we study the four atoms forming a square.\ The $l$ $=$ $2$ subradiant mode, possessing the lowest decay rate, is more narrowly directional than the other ones of $l$ $=$ $1$ and $3$, which have the same $\Omega_f(\theta,\phi)$.\ Furthermore, in this specific structure, two polarizations of light excitations generate the same pattern of the far-field scattering with the rotational symmetry of $\phi$ $\rightarrow$ $\phi$ $+$ $\pi/2$.\ This applies to all the number of atoms $N$ $=$ $4n$ with integers $n$, which therefore preserves the $C_4$ rotational symmetry in their scattering properties.\ This can be also seen from the linearly polarized light we use here, which rotates the dipole moment by $\pi/2$ in $\phi$.\ We note that the far-field scatterings of the $l$ $=$ $1$ and $3$ modes are the same due to the symmetry of $l$ $\rightarrow$ $-l$, which will be further explained in the end of this subsection. 

\begin{figure}[t]
\centering
\includegraphics[width=10.0cm,height=5.0cm]{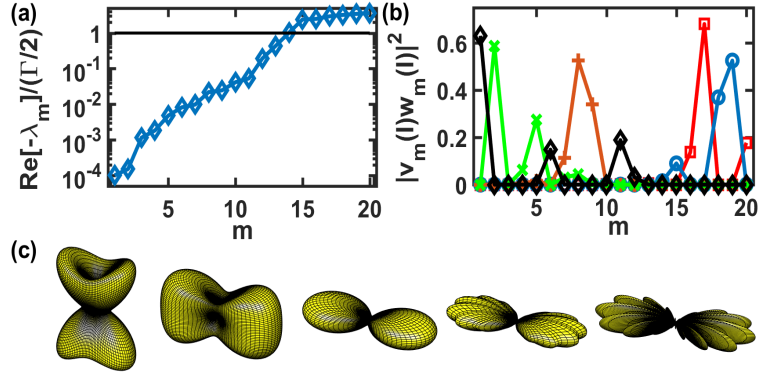}
\caption{The decay rates, normalized weightings, and $\Omega_f(\theta,\phi)$ of a single ring, $N=20$ with $r/\lambda$ $=$ $0.5$.\ (a) The spontaneous decay rates are numerically derived from the real part of the eigenvalues $\lambda_m$ and are shown in logarithmic scale with an ascending order.\ (b) The normalized weightings on the eigenmodes for $l$ $=$ $1$, $2$, $5$, $9$, and $10$ ($\bigcirc$, $\square$, $+$, $\times$, and $\diamond$ respectively).\ Corresponding far-field $\Omega_f(\theta,\phi)$ are plotted horizontally in (c), in the same viewing angle of Figure \ref{fig2}.}\label{fig4}
\end{figure}

For many atoms on a single ring, we use $N$ $=$ $20$ as an example in Figure \ref{fig4}.\ For this $N$, $C_4$ rotational symmetry sustains, and therefore, $\Omega_f(\theta,\phi)$ are the same for two linear polarizations.\ In this configuration, the lowest subradiant eigenmode has $10^{-4}$ times of the natural decay rate, as shown in Figure \ref{fig4}(a).\ This extremely small scale of the decay rate can be further reduced as $r$ decreases.\ We select some of the HPI states in Figure \ref{fig4}(b), which occupy several eigenmodes with significant weightings migrating from the superradiant ($l$ $=$ $1$ and $2$) to the subradiant ones ($l$ $=$ $5$, $9$, and $10$).\ The HPI super- and subradiant states can be approximately determined and distinguished by locating the eigenmode of $m$ $=$ $14$ in Figure \ref{fig4}(a), which is about to pass below the line of $\lambda_m$ $=$ $-\Gamma/2$.\ In Figure \ref{fig4}(c), the $\Omega_f(\theta,\phi)$ is plotted accordingly for the selected modes in \ref{fig4}(b).\ Except for the superradiant mode of $l$ $=$ $N$, which has a clear forward-backward scattering similar to the previous cases of few atoms, the other superradiant modes of $l$ $=$ $1$ and $2$, for example, also show an oblique and side scatterings respectively.\ As $l$ increases toward the most subradiant modes ($l$ $=$ $10$), a clear side scattering at the right angles emerges, and this most subradiant HPI state further narrows $\Omega_f(\theta,\phi)$ and becomes directional.\ 

We note of the identical $\Omega_f(\theta,\phi)$ with $l$ $\rightarrow$ $-l$ (or $N$ $-$ $l$) for even number of the atoms in general, and for odd number of atoms only when $r$ $\rightarrow$ $0$.\ For small $r$, the far-field property $\Omega_f(\theta,\phi)$ of Eq. (\ref{far}) can be approximately reduced to the sum of imprinted helical phases $e^{i2l\pi(\beta_\phi-\alpha_\phi)/N_\phi}$.\ This gives in general a sum of cosine functions without the detail spatial phases from the atomic distributions of $\r_{\alpha\beta}$, therefore $\Omega_f(\theta,\phi)$ is the same for $l$th and ($N$ $-$ $l$)th HPI states.\ For a finite $r$, only even number of the atoms $N$ sustains the symmetry of $l$ $\rightarrow$ $-l$ in $\Omega_f(\theta,\phi)$.\ This can be seen from the pairwise and spatial phase contributions of $\k_R\cdot\r_{\alpha\beta}$.\ These include a combination of $C^N_2$ cosine functions with $N$ nearest-neighbor components ($\alpha$ $=$ $\beta$ $+$ $1$) and with $N(N-3)/2$ diagonals in the geometry of $N$-polygon.\ The nearest-neighbor components pair up and interchange with $l$ and $-l$.\ For the diagonals, they can be further grouped into $(N/2$ $-$ $2)$ different lengths (next nearest-neighbor, next next nearest-neighbor, etc.) with $N$ components respectively, and the diagonal with the maximal length ($2r$) with $N/2$ components.\ Again the $N$ components in the respective groups can be interchanged with $l$ and $-l$.\ The $N/2$ components in the maximal diagonal go back to themselves as $l$ $\rightarrow$ $-l$.\ This is due to the form of $\cos(l\pi+\k_R\cdot\r_{\alpha\beta})$ which is the same as $\cos(-l\pi+\k_R\cdot\r_{\alpha\beta})$ with a phase difference of $2l\pi$.

\subsection{Stacked rings.} 

Moreover, we investigate the far-field scattering properties of the stacked rings.\ When more atoms are involved, the atomic coherences can be built up to enhance the cooperative emissions in either the super or subradiant decay rates, the cooperative frequency shift, and the directionality of the scattering.\

\begin{figure}[t]
\centering
\includegraphics[width=10.0cm,height=5.0cm]{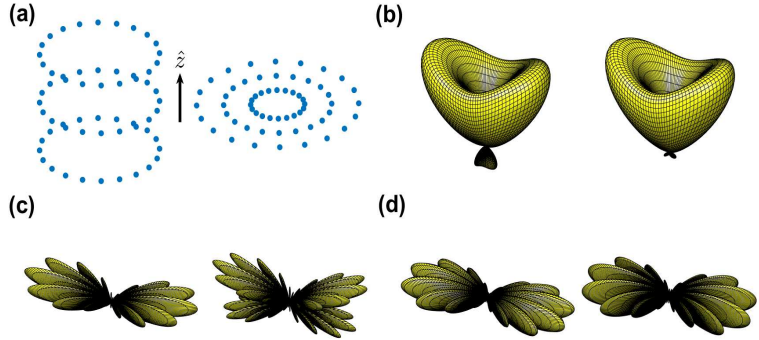}
\caption{Schematic stacked ring arrays and their far-field properties $\Omega_f(\theta,\phi)$ with $r/\lambda$ $=$ $0.5$.\ For the stacked ring arrays with $N_\phi$ $=$ $20$ in (a), $\Omega_f(\theta,\phi)$'s of the two and three $\hat z$-stacked rings ($d_z/\lambda$ $=$ $0.35$) of the HPI states with (b) $l$ $=$ $1$ and (c) $10$ indicate a forward scattering with more rings.\ (d) Far-field scattering of the HPI state with $l$ $=$ $9$ for two and three $\hat r$-stacked rings show a narrowed scattering, compared to Figure \ref{fig4}(c).\ The viewing angles are the same as Figure \ref{fig2}.}\label{fig5}
\end{figure}

In Figure \ref{fig5}, we investigate two basic types of the stacked rings, $\hat z$- and $\hat r$-stacked rings, which are integrated along the excitation and radial directions respectively.\ Figure \ref{fig5}(a) shows schematically three $\hat z$- and $\hat r$-stacked rings, where an extra parameter of $d_z$ characterizes the distance between the rings in the former type, and we consider only concentric rings in the latter.\ For the superradiant mode of HPI state with $l$ $=$ $1$ in Figure \ref{fig5}(b), the forward scattering is enhanced from two to three stacked rings, breaking the symmetry of forward-backward scattering in Figure \ref{fig4}(c).\ As we vary $d_z$, the relative strengths between the forward and backward scatterings can vary, which however are always larger than one when $d_z$ $\lesssim$ $\lambda$.\ Therefore, the far-field scattering is hugely influenced by the interference between the rings. 

Similarly for the subradiant mode in Figure \ref{fig5}(c), an oblique scattering toward the forward and backward directions are reinforced as more atoms are integrated in the $\hat z$-stacked rings.\ As we stack up more rings, the side scattering at the right angle in Figure \ref{fig4}(c) can reappear as well.\ Whether the atomic system supports an oblique or side scattering depends on the $r$, $d_z$, and the number of the stacked rings, which again indicates the interplay of the phase interferences within and between each rings.\ 

\begin{figure}[t]
\centering
\includegraphics[width=10.0cm,height=5.0cm]{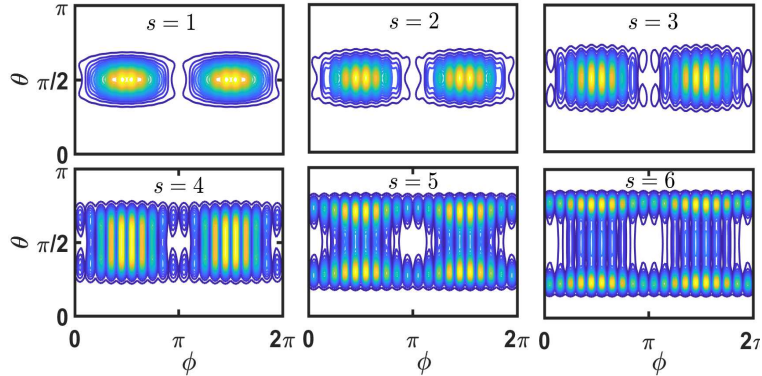}
\caption{Far-field scattering for $\hat r$-stacked rings.\ The scattering intensity shows narrowing in $\phi$ direction as $s$ increases, and further in $\theta$ direction as well at $s$ $=$ $5$ and $6$, indicating a crossover from the sub- to the superradiant mode of HPI state with $l$ $=$ $9$.\ Three-dimensional plots of the cases of $s$ $=$ $1$, $2$, $3$ are shown in Figure \ref{fig4}(c) and Figure \ref{fig5}(d) correspondingly.}\label{fig6}
\end{figure}

In the last example of Figure \ref{fig5}(d), we study the HPI state with $l$ $=$ $9$ in the $\hat r$-stacked rings.\ Compared to this subradiant mode in a single ring in Figure \ref{fig4}(c), we find a narrowing effect on the far-field scattering in $\phi$ direction.\ As we put more rings together (a total number of $s$ rings), passing the range of $sr$ $=$ $1.5\lambda$ with $s$ $=$ $3$, the far-field scattering toward the forward and backward directions starts to appear, which we further demonstrate in Figure \ref{fig6}.\ With $s$ $=$ $4$, we see a broadening in $\theta$ direction, which indicates a crossover from the sub- to the superradiant mode for the chosen $l$.\ The superradiant modes at $s$ $=$ $5$ and $6$ further show the narrowing effects in both $\theta$ and $\phi$ directions on the far-field property, in huge contrast to the purely forward and backward scattering along $\hat z$ for $l$ $=$ $N_\phi$.\ We note that the number of the narrowed peaks depends on the number of the atoms on the ring, which is $N_\phi$ $=$ $20$ in Figure \ref{fig6}.\ This is also a signature of the HPI states in equation (\ref{DM2}) with $N_\phi$ atoms imprinted by OAM.\ The other modes of $l$ $=$ $8$ and $10$ also have this signature for the $\hat r$-stacked rings.\ For the $\hat z$-stacked rings, a clear feature of $N_\phi$-periodic interference only manifests in the $l$ $=$ $N_\phi/2$ states, as shown in Figures \ref{fig4}(c) and \ref{fig5}(c).\

\subsection{Circular polarization.}

\begin{figure}[t]
\centering
\includegraphics[width=10.0cm,height=5.0cm]{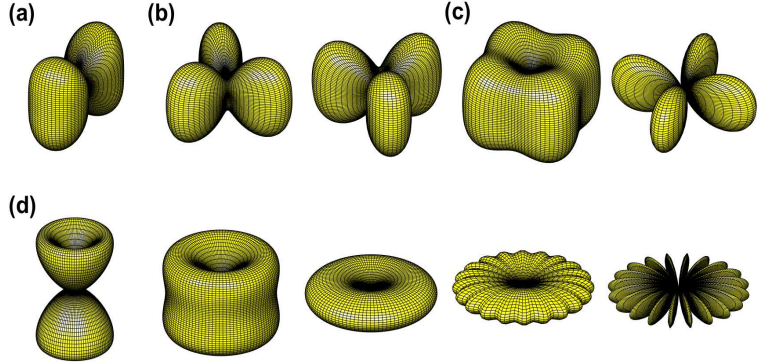}
\caption{Far-field scattering from circularly polarized excitations.\ The far-field scattering properties of the HPI subradiant states are shown for (a) two-atom ($l$ $=$ $1$), (b) three-atom ($l$ $=$ $1$ and $2$), and (c) four-atom cases ($l$ $=$ $1$ and $2$), at $r/\lambda$ $=$ $0.2$.\ (d) The far-field scatterings of a single ring (same $N$ and $r$ of Figure \ref{fig4}) with a circularly polarized excitation for HPI states of $l$ $=$ $1$, $2$, $5$, $9$, and $10$ are illustrated horizontally.}\label{fig7}
\end{figure}

For circularly polarized excitations [$(\hat x\pm \hat y)/\sqrt{2}$] of HPI states, we expect of more symmetric scattering patterns than linear polarizations.\ In Figure \ref{fig7}, we choose HPI subradiant states with $l$ $=$ $1$ or $2$ for the few atoms case, and for many atoms, we compare with a single ring structure in Figure \ref{fig2}.\ For any uniform polarizations in general, $(\hat \R\cdot\hat \p)^2$ in equation (\ref{far}) can be replaced by $|\hat \R\cdot\hat \p|^2$, and as such, the handedness of the circular polarization does not matter to the far-field property.\ For the two-atom case, the scattering property can be derived by substituting the prefactor of equation (\ref{far2}), $\sin^2\theta\cos^2\phi$, by $\sin^2\theta/2$.\ Similar to the linear polarizations in Figure \ref{fig2}, vanishing scattering intensity resides on $\hat y$-$\hat z$ plane, while in contrast two polar angles of $\pi/2$ $\pm$ $\theta_m$ on the $\hat x$-$\hat z$ plane can be identified at the maximal scattering in Figure \ref{fig7}(a).\ This is exactly the average of the far-field scattering of HPI state with $l$ $=$ $1$ in Figure \ref{fig2}.\ For three and four atoms, the subradiant modes have scattering peaks at discrete azimuthal angles $\phi$ $=$ $\phi_s$ $+$ $2\pi/N$, preserving a discrete rotational symmetry $C_N$ of $\phi$.\ $\phi_s$ is the offset of the angle, depending on which subradiant mode we consider in Figures \ref{fig7}(b) and \ref{fig7}(c).

For a single ring with many atoms, we compare Figure \ref{fig7}(d) to Figure \ref{fig4}(c), and similarly the far-field scattering goes toward the transverse direction from the super- to the subradiant modes.\ In addition, the circularly polarized excitation preserves the azimuthal symmetry in the super- ($l$ $=$ $1$ and $2$) and subradiant ($l$ $=$ $5$) scattering patterns.\ For even more subradiant modes ($l$ $=$ $9$ and $10$), the rotational symmetry $C_N$ of $\phi$ emerges and contrasts with simply narrowing peaks in Figures \ref{fig4}, \ref{fig5}, and \ref{fig6}.\ Similarly in the stacked rings along $\hat z$ or $\hat r$, circularly polarized excitation further endows the scattering patterns with the azimuthal and discrete rotational symmetries respectively to the super- and subradiant modes in Figure \ref{fig5}.\ The narrowing effect in the $\hat r$-stacked rings also appears to circular polarizations as in Figure \ref{fig6} with increasing $s$, and again with an additional $C_N$ symmetry.

\subsection{Radial and azimuthal polarizations.}

Other than uniformly polarized excitations, finally we study the scattering properties from the radially and azimuthally polarized excitations which can be generated and tailored \cite{Du2017} to manipulate spatially-dependent dipole orientations.\ The dipole-dipole interactions in equations (\ref{F}) and (\ref{G}) can be straightforwardly generalized by changing $[$$1$ $-$ $(\hat \p\cdot \hat r_{\mu\nu})^2$$]$ to $[$$(\hat \p_\beta^*\cdot\hat \p_\alpha)$ $-$ $(\hat \p_\beta^*\cdot\hat r_{\alpha\beta})$ $(\hat \p_\alpha\cdot\hat r_{\alpha\beta})$$]$ and correspondingly in the term of $[$$1$ $-$ $3(\hat \p\cdot \hat r_{\mu\nu})^2$$]$.\ Similarly, the far-field property of equations (\ref{intensity}) and (\ref{far}) can be also generalized by replacing $[$$1$ $-$ $(\hat\R\cdot\hat \p)^2$$]$ with $[$$(\hat \p_\beta^*\cdot\hat \p_\alpha)$ $-$ $(\hat \p_\beta^*\cdot\hat \R)$ $(\hat \p_\alpha\cdot\hat\R)$$]$ and moving it into the double sums $\sum_{\alpha,\beta}$.

\begin{figure}[t]
\centering
\includegraphics[width=10.0cm,height=5.0cm]{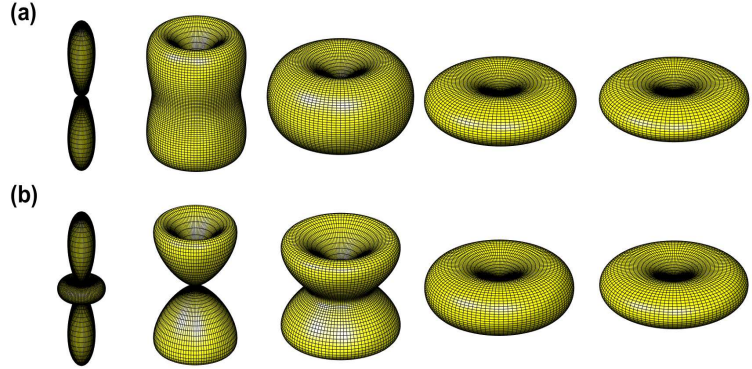}
\caption{Far-field scattering from radially and azimuthally polarized excitations.\ The far-field scattering properties of the HPI subradiant states are plotted for a single ring with the same $N$ and $r$ of Figure \ref{fig4}, which are excited by (a) radial and (b) azimuthal polarizations.\ The modes of $l$ $=$ $1$, $2$, $3$, $5$, and $6$ are illustrated horizontally.}\label{fig8}
\end{figure}

The radial and azimuthal polarizations for the dipole can be denoted as $\hat \p_\alpha$ $=$ $\hat e_r$ and $\hat e_\phi$ which are respectively [$\cos\phi(\alpha)\hat x$ $+$ $\sin\phi(\alpha)\hat y$] and [$-\sin\phi(\alpha)\hat x$ $+$ $\cos\phi(\alpha)\hat y$].\ The dipole orientation rotates with the angles $\phi(\alpha)$ $=$ $2\pi\alpha_\phi/N_{\phi}$ which amounts to the imprinted phase via spin polarizations.\ Unlike the circular polarizations which have definite spin angular momentum ($\pm 1\hbar$) in a photon, the radial and azimuthal polarizations can only relate to the circular ones by $\hat e_r \pm i\hat e_\phi$ $=$ $(\hat x\pm i\hat y)e^{\mp i\phi}$.\ This indicates that an equal superposition with $\pi/2$ phase shift between the radial and azimuthal polarizations of light is the same as the left (right)-handed circular polarization with one quanta shift ($\mp 1\hbar$) of OAM.\ In contrast to this simple relation between the light polarizations, the scattered polarization of the electric field has both contributions from the excited dipoles and the observer, $\propto$ [$\hat \p_\alpha$ $-$ $\hat\R(\hat \p_\alpha\cdot\hat\R)$], making the resulting far-field scattering property lack of this direct relation.\ Nevertheless, the mode shift ($l$ $\rightarrow$ $l$ $+$ $1$) for small $l$ in the scattering patterns can still happen.\ In Figure \ref{fig8}, we demonstrate the far-field scattering patterns from radially and azimuthally polarized excitations.\ Both $l$ $=$ $1$ HPI states resemble the superradiant ($l$ $=$ $N_\phi$ or equivalently $l$ $=$ $0$) scattering patterns characterized by the forward and backward directions.\ As $l$ increases, we find that the far-field scattering of $l$ $=$ $2$ excited by the azimuthal polarization has the oblique structure of $l$ $=$ $1$ mode by the circular polarization in Figure \ref{fig7}(d).\ In addition, the $l$ $=$ $2$ mode by the circular polarization is analogous to the average of radially and azimuthally polarized mode patterns ($l$ $=$ $3$).\ When $l$ $\approx$ $N_\phi/2$, we find that all of the patterns from circularly, radially, or azimuthally polarized excitations approach to each other.\ This is due to both the symmetry of $l$ $\rightarrow$ $-l$ in $\Omega_f(\theta,\phi)$ and the discrete rotational symmetry $C_N$ are satisfied.\ Similarly, the stacked rings along $\hat z$ or $\hat r$ respectively enhances the forward scattering and the narrowing effect as in Figure \ref{fig5} with again an additional $C_N$ symmetry.

\section{Discussion}

For the source of light with OAM, we use the paraxial approximation for the LG modes \cite{Allen1992, Barnett1994, Cerjan2011}, which can be satisfied when $f$ $=$ $\lambda/(2\pi w_0)$ (beam waist $w_0$) is much smaller than one.\ Under this condition, the extra spatially-dependent phase [$e^{-ikr^2/(2R(z))}$] from the radius of curvature [$R(z)$] of the wavefront vanishes at $z$ $=$ $0$ due to infinite $R(z)$.\ Therefore, the HPI state can be genuinely prepared with the phase of $e^{il\phi}$ when interacting with the photons with the $l$th OAM.\ However, the longitudinal polarization can be generated due to LG modes gradient along the direction of the linear polarization \cite{Allen1992, Barnett1994, Cerjan2011}, which reduces the light-matter coupling efficiency.\ Taking the advantage of precise atomic spatial manipulations and tightly-focused beams \cite{Du2017}, the HPI state preparation can be optimized to overcome this inefficiency.\   

Our proposed scheme to manipulate the HPI states can be limited by how large the OAM of light is available for the scalable atomic system.\ A recent advancement shows that as high as more than $10,000$ OAM can be generated \cite{Fickler2016}, thus making our scheme flexible and pragmatic enough up to several thousands of atoms.\ To have a strong coupling in the atom-atom interactions, the requirement is more stringent in gaseous systems since the condition of $r$ $\lesssim$ $\lambda$ already reaches Bose-Einstein condensation.\ On the contrary, it is more suitable in the systems of artificial atoms \cite{Hanson2008, Buluta2011} using the superconducting qubits or silicon-vacancy color centers \cite{Sipahigil2016}.\ Respectively, $\mu$m scale of superconducting circuits are driven by microwave fields (several tens of GHz corresponding to $\lambda$ $\sim$ $30$ $\mu$m), while the color centers can be manipulated with $40$-nm precision driven by an infrared light ($400$ THz corresponding to $\lambda$ $=$ $750$ nm).\ Both systems allow for strong dipole-dipole interactions ($r/\lambda$ $\sim$ $0.05$) with high controllability of the atomic distributions, which are therefore promising to implement the HPI state engineering of super- or subradiant states.\ Furthermore, our scheme can be favorable to store and manipulate the quantum information using OAM of light which in principle has infinite capacity of entanglement, and is potentially useful in quantum computing with high-dimensional quantum gates \cite{Babazadeh2017}.

Finally, the far-field scattering properties here offer distinguished fingerprints that can be traced back to the atomic spatial distributions and polarization configurations.\ Therefore, the scattered patterns not only provide useful information for light collections, but incorporate rich details of light-matter interactions.\ Our HPI states are one example of systematically delineating the full spectra of scattered mode patterns.\ As the development of controlling atomic positions progresses, we expect of more collective phenomena to emerge in addition to light reflections \cite{Shahmoon2017} or many-body subradiant state excitations \cite{Facchinetti2016, Plankensteiner2017}.
\appendix
\section{Lindblad form of dissipation with resonant dipole-dipole interactions.}  
The theoretical analysis for the fluorescence and light scattering is based on the Lindblad forms of the spontaneous emissions.\ The general spontaneous emission process involves the long-range dipole-dipole interaction \cite{Stephen1964,Lehmberg1970}.\ This pairwise interaction originates from the rescattering events in the common quantized light field.\ For an arbitrary quantum operator $\hat{Q}$, the Heisenberg equation in a Lindblad form gives 
\bea
\frac{d\hat{Q}}{dt} = -i\sum_{\mu\neq\nu}^N\sum_{\nu=1}^N G_{\mu\nu}[\hat{Q},\hat{\sigma}_\mu^+\hat{\sigma}_\nu^-] + \mathcal{L}_s[\hat{Q}],\label{Q}
\eea
where for the spontaneous emission,
\bea
\mathcal{L}_s[\hat{Q}]=-\sum_{\mu,\nu=1}^N\frac{F_{\mu\nu}}{2}\left(\hat{\sigma}_\mu^+\hat{\sigma}_\nu^-\hat{Q}+\hat{Q}\hat{\sigma}_\mu^+\hat{\sigma}_\nu^- -2\hat{\sigma}_\mu^+\hat{Q}\hat{\sigma}_\nu^-\right).
\eea
The dipole operator is $\hat{\sigma}_\mu^-$ ($\hat{\sigma}_\mu^+$) where $\hat{\sigma}_\mu^-$ $\equiv$ $|g\rangle_\mu\langle e|$ and $\hat{\sigma}_\mu^-$ $\equiv$ $(\hat{\sigma}_\mu^+)^\dag$.\ The pairwise frequency shift $G_{\mu\nu}$ and decay rate $F_{\mu\nu}$ are \cite{Lehmberg1970}
\bea
F_{\mu\nu}(\xi)\equiv&&
\frac{3\Gamma}{2}\bigg\{\left[1-(\hat\p\cdot\hat{r}_{\mu\nu})^2\right]\frac{\sin\xi}{\xi}
\nonumber\\
&&+\left[1-3(\hat\p\cdot\hat{r}_{\mu\nu})^2\right]\left(\frac{\cos\xi}{\xi^2}-\frac{\sin\xi}{\xi^3}\right)\bigg\},\label{F}\\
G_{\mu\nu}(\xi)\equiv&&\frac{3\Gamma}{4}\bigg\{-\Big[1-(\hat\p\cdot\hat{r}_{\mu\nu})^2\Big]\frac{\cos\xi}{\xi}
\nonumber\\&&
+\Big[1-3(\hat\p\cdot\hat{r}_{\mu\nu})^2\Big]
\left(\frac{\sin\xi}{\xi^2}+\frac{\cos\xi}{\xi^3}\right)\bigg\}\label{G}, 
\eea
where $\Gamma$ is the single-particle natural decay rate of the excited state, $\xi$ $=$ $|\k_L| r_{\mu\nu}$, and the interparticle distance $r_{\mu\nu}$ $=$ $|\mathbf{r}_\mu-\mathbf{r}_\nu|$.\ The above expressions are valid for uniformly polarized excitations of the dipole orientations $\p$.

The time evolutions of the HPI states can be determined by the eigenvalues and eigenvectors of the coupling matrix $\hat M$ with $M_{\mu\nu}$ $=$ $(-F_{\mu\nu}+i2G_{\mu\nu}\delta_{\mu\neq \nu})e^{-i\k_L\cdot(\r_\mu-\r_\nu)}/2$ in the bare state bases $|\psi_\mu\rangle$.\ Denote the eigenvalues and eigenvectors as $\lambda_m$ and $\hat U$ respectively, the time evolution of the HPI state $|\Psi(t)\rangle$ $=$ $h_l(t)|\Phi_l\rangle_{\rm HPI}$ reads \cite{Jen2016_SR,Jen2016_SR2,Jen2017_MP}
\bea
h_l(t)=&&\sum_{m=1}^N v_m(l)e^{\lambda_m t}w_m(l),~
v_m(l)=\sum_{\mu_z=1}^{N_z}\sum_{\mu_\phi=1}^{N_\phi}\frac{U_{\mu m}}{\sqrt{N}}e^{-i\frac{2l\pi}{N_\phi}(\mu_\phi-1)},\nonumber\\
w_m(l)=&&\sum_{\nu_z=1}^{N_z}\sum_{\nu_\phi=1}^{N_\phi}\frac{U_{m\nu}^{-1}}{\sqrt{N}}e^{i\frac{2l\pi}{N_\phi}(\nu_\phi-1)},\label{evolve}
\eea
where the atomic position index $\nu$ is implicitly $(\nu_z$$-$$1)$$N_\phi$ $+$ $\nu_\phi$, which is the same as $\mu$.\ The eigen-spectrum of $\lambda_m$ involves both super- and subradiant decay rates along with the associated frequency shifts.\ $|v_m(l)|^2$ is essentially the fidelity of $|\Phi_l\rangle_{\rm HPI}$ to the $m$th eigenstate while $|v_m(l)w_m(l)|^2$ gives a measure of how much $\lambda_m$ contributes to the HPI state dynamics. 

\section{Far-field scattering.}  
The far-field scattering properties provide extra information in characterizing the HPI states and the atomic system.\ Similar ring lattice has been used to prepare Rydberg states \cite{Olmos2010} which show collective effects in the photon emissions.\ Here we use the general expression of the far-field scattering from the two-level atoms \cite{Lehmberg1970},
\bea
\left\langle\vec E^*(\R,t')\vec E(\R,t)\right\rangle=&&\left(\frac{\omega_{eg}^2}{4\pi\epsilon_0 c^2}\right)^2\frac{|\vec p|^2}{R^2}\left[1-(\hat \R\cdot \hat \p)^2\right]\nonumber\\
&&\times\sum_{\alpha,\beta=1}^N e^{i\k_R\cdot\r_{\alpha\beta}}\left\langle\hat{\sigma}_\alpha^+(t')\hat{\sigma}_\beta^-(t)\right\rangle,\label{intensity}
\eea
where $\omega_{eg}$ is the energy difference, $R$ $=$ $|\R|$, $\r_{\alpha\beta}$ $=$ $\r_\alpha$ $-$ $\r_\beta$, and the orientation of the dipole moment $\vec p$ is determined by the polarization of the excitation.\ The far-field derivation assumes that the observation point is much farther from the atoms, such that $\omega_{eg}|\R-\r_\alpha|/c$ $\gg$ $1$.\ This also suggests that the radiation mode $\k$ $\approx$ $\k_R$ $\parallelsum$ $\R_\alpha$ [$=$ $(\R-\r_\alpha)$] in Figure \ref{fig1}, which indicates of the retarded phase $e^{ik_R(R_\alpha-R)}$ $\approx$ $e^{-i\k_R\cdot\r_\alpha}$.\ Similar and more general expression can be also derived for a four-level atomic system \cite{James1993} (three Zeeman levels in the $J$ $=$ $1$ excited state, as in strontium atoms), which takes Eq. (\ref{intensity}) in a tensor form of dipole transitions.\

At equal time of Eq. (\ref{intensity}), we obtain the radiation field intensity.\ Put the HPI states $|\Psi(t)\rangle$ into Eq. (\ref{intensity}) in Schr\"{o}dinger picture, we have
\bea
\frac{\left\langle\vec E^*(\R,t)\vec E(\R,t)\right\rangle}{I_0(t)}=&&\left[1-(\hat \R\cdot \hat \p)^2\right]\sum_{\alpha,\beta=1}^N e^{i(\k_R-\k_L)\cdot\r_{\alpha\beta}} \nonumber\\
&&\times\frac{1}{N}e^{\frac{i2l\pi}{N_\phi}(\beta_\phi-\alpha_\phi)},\label{far}
\eea
where $I_0(t)$ $=$ $I_nh_l^*(t)h_l(t)$ is the time-evolving fluorescence intensity with $I_n$ $\equiv$ $(\omega_{eg}|\vec p|)^2/(4\pi\epsilon_0 c^2R)^2$, and again $\alpha(\beta)$ has an implicit dependence of $\alpha_\phi(\beta_\phi)$.\ Equation. (\ref{far}) characterizes the far-field scattering property from the HPI states prepared by an excitation field of $l$th OAM, which involves the interplay of the atomic distributions $\r_{\alpha\beta}$ and the imprinted phases $e^{i2l\pi/N_\phi}$.

\ack
This work is supported by the Ministry of Science and Technology (MOST), Taiwan, under the Grant No. MOST-104-2112-M-001-006-MY3 and No. MOST-106-2112-M-001-005-MY3.\ H.H.J is partially supported by a grant from MOST of No. 106-2811-M-001-130, as an assistant research scholar in IOP, Academia Sinica.\ We are grateful for the support of NCTS ECP1 (Experimental Collaboration Program) and for stimulating discussions with Dr. Wan-Ju Li on $C_4$ rotational symmetry in the scattering properties of the single ring geometry.

\end{document}